\documentclass[3p, twocolumn, number]{elsarticle}
\usepackage[T1]{fontenc}
\usepackage[utf8]{inputenc}
\usepackage{color}
\usepackage{textcomp}
\usepackage{amsthm}
\usepackage{amsmath}
\usepackage{graphicx}
\usepackage{esint}
\usepackage[unicode=true,
 bookmarks=false,
 breaklinks=false,pdfborder={0 0 0},backref=false,colorlinks=true]
 {hyperref}
\hypersetup{pdftitle={Free energy of the bcc-liquid interface and the Wulff shape as predicted by the Phase-Field Crystal model},
 pdfauthor={Frigyes Podmaniczky, Gyula I. Tóth, Tamás Pusztai, László Gránásy},
 pdfsubject={physics},
 pdfkeywords={Solidification, Surface energy anisotropy, Wulff shape, Phase-Field Crystal model}}
\usepackage{breakurl}

\makeatletter

\providecommand{\tabularnewline}{\\}
\newcommand{\lyxdot}{.}

\journal{Journal of Crystal Growth}
\renewcommand{\citet}{\cite}

\makeatother

\begin{document}

\title{Free energy of the bcc-liquid interface and the Wulff shape as predicted
by the Phase-Field Crystal model}

\author[loc-bp]{Frigyes Podmaniczky}

\author[loc-bp]{Gyula I. Tóth}

\author[loc-bp]{Tamás Pusztai}

\author[loc-bp,loc-brunel]{László Gránásy\corref{cor-author}}

\ead{Granasy.Laszlo@wigner.mta.hu}

\cortext[cor-author]{Corresponding author.}

\address[loc-bp]{Institute for Solid State Physics and Optics, Wigner Research Centre
for Physics, P O Box 49, H-1525 Budapest, Hungary}

\address[loc-brunel]{BCAST, Brunel University, Uxbridge, Middlesex, UB8 3PH, UK}
\begin{abstract}
The Euler-Lagrange equation of the phase-field crystal (PFC) model
has been solved under appropriate boundary conditions to obtain the
equilibrium free energy of the body centered cubic crystal-liquid
interface for 18 orientations at various reduced temperatures in the
range $\epsilon\in\left[0,0.5\right]$. While the maximum free energy
corresponds to the $\left\{ 100\right\} $ orientation for all $\epsilon$
values, the minimum is realized by the $\left\{ 111\right\} $ direction
for small $\epsilon\,(<0.13)$, and by the $\left\{ 211\right\} $
orientation for higher $\epsilon$. The predicted dependence on the
reduced temperature is consistent with the respective mean field critical
exponent. The results are fitted with an eight-term Kubic harmonic
series, and are used to create stereographic plots displaying the
anisotropy of the interface free energy. We have also derived the
corresponding Wulff shapes that vary with increasing $\epsilon$ from
sphere to a polyhedral form that differs from the rhombo-dodecahedron
obtained previously by growing a bcc seed until reaching equilibrium
with the remaining liquid. \end{abstract}
\maketitle

\section{Introduction}

The anisotropy of the crystal-liquid interface free energy ($\gamma_{hkl}$)
reflects differences in the interface structure for different orientations,
and may play an essential role in determining the morphology of growing
crystals \citet{key-1}. Anisotropy is needed for dendritic structures,
and apparently a detailed knowledge on anisotropy is required to fully
understand the growth morphology \citet{key-2}. Experimentally, the
anisotropy of the interface free energy can be deduced from the shape
of liquid inclusions in a solid matrix (though, one needs to be careful
to relax all the stresses before converting the shape into anisotropy)
\citet{key-3,key-4,key-5,key-6}. Anisotropy has also been evaluated
on the basis of the assumption that the dendrite growth directions
correspond to the maximum stiffness, and minimizing the deviation
between the calculated minima of an appropriately parameterized interface
stiffness function and the growth directions of dendrites found in
thin coatings experimentally \citet{key-7}. Other methods evaluate
the interface free energy and its anisotropy from molecular dynamics
simulations \citet{key-8,key-9,key-10,key-11,key-12} using empirical
model potentials such as the embedded atom potential. Whether experiment
\citet{key-5,key-6,key-7} or atomistic simulation \citet{key-8,key-9,key-10,key-11,key-12},
the anisotropic interface free energy data are usually fitted by the
cubic harmonic expansion series introduced by Fehlner and Vosko \citet{key-13}.
Often only a few low index orientations are considered (typically
$\left\{ 100\right\} $, $\left\{ 110\right\} $, and $\left\{ 111\right\} $),
and a second-order cubic harmonic expansion is employed \citet{key-5,key-6,key-7,key-8,key-9}.

Theoretical predictions for the anisotropy of the crystal-liquid interface
in 3D emerge mostly from the early broken-bond models for the fcc,
bcc, hcp, and dc structures \citet{key-14,key-15,key-16} (utilizing
former results for the crystal-vapor interfaces \citet{key-17,key-18,key-19,key-20,key-21}),
from the classical density functional theory \citet{key-22,key-23},
and recently for the fcc and bcc structures from the Phase-Field Crystal
(PFC) approach \citet{key-23,key-24} (a simple dynamical density
functional theory \citet{key-25,key-26,key-27}). Some analytical
predictions based on the approximations of the PFC model are also
available: A multi-scale analysis has been used by Wu and Karma \citet{key-28}
to evaluate the anisotropy of the interfacial fee energy near the
critical point. They have approximated the equation of motion of the
PFC model by a set of coupled equations that describe the time evolution
of the amplitudes of the dominant density waves. Analyzing the stationary
solution, they have concluded that the anisotropy is independent of
the reduced temperature, a finding that accords with the results of
Majaniemi and Provatas \citet{key-29}, who have used the local volume
averaging method to obtain amplitude equations for liquid–solid interfaces
that are broad relative to the periodicity of the crystalline phase.
In these studies, the independence of the anisotropy from the reduced
temperature follows from approximations, which lead to weakly fourth-order
amplitude theories of the Ginzburg-Landau type, from which all material
parameters can be scaled out \citet{key-28,key-29}. As a result,
the anisotropy of the solid-liquid interface free energy depends only
on the crystal structure. This independence of the anisotropy from
the reduced temperature is, however, unphysical, as the anisotropy
must vanish, when the correlation length (the width of the solid-liquid
interface) diverges at the critical point, as indeed suggested by
the equilibrium shapes of finite size clusters from 2D PFC simulations
\citet{key-30,key-31,key-32}. It is, however, important to test the
anisotropy of the solid-liquid interface free energy by equilibrium
simulations for the flat interface in 3D, which is free from size
effects, since the latter is known to influence the equilibrium shape
considerably \citet{key-21,key-30}.

Accordingly, in this paper, we are going to demonstrate that the free
energy of the flat bcc-liquid interface depends not only on the orientation
but also on the reduced temperature. We map out the orientation dependence
in detail at several reduced temperatures, then fit the results with
an expression based on an eight-term Kubic harmonic expansion \citet{key-13,key-aux-1,key-aux-2}
to represent the orientation dependence in a closed form, and determine
the respective Wulff shapes, a mathematical construction \citet{key-17,key-33}
to which the equilibrium shape, minimizing the interfacial contribution
to the cluster free energy for a given volume, tends for large particle
sizes.

\begin{figure*}
\hfill{}%
\begin{tabular}{ll}
{\footnotesize (a)} & {\footnotesize (b)}\tabularnewline
\includegraphics[scale=0.25]{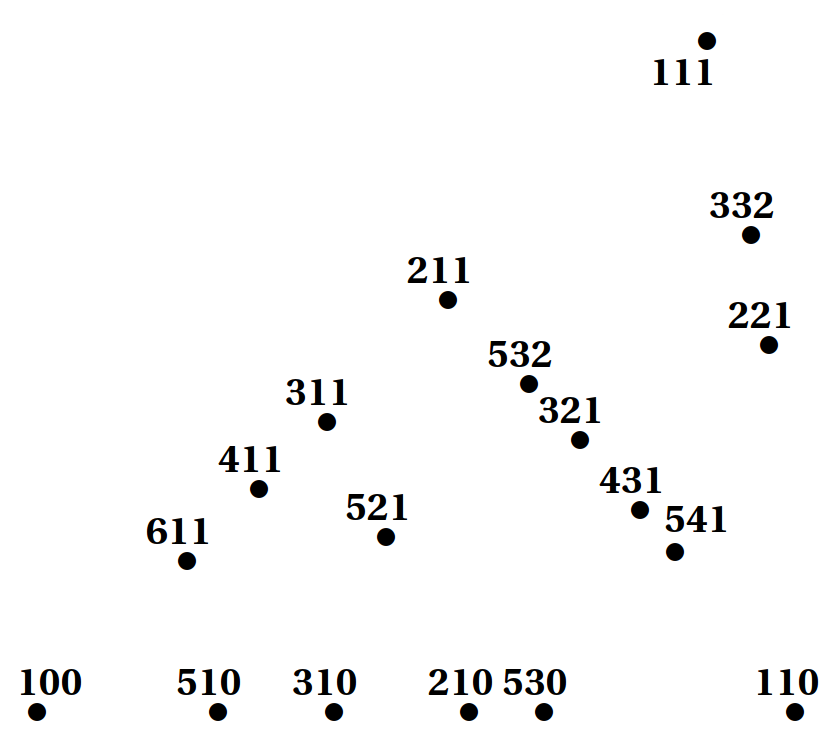} & \includegraphics[scale=0.25]{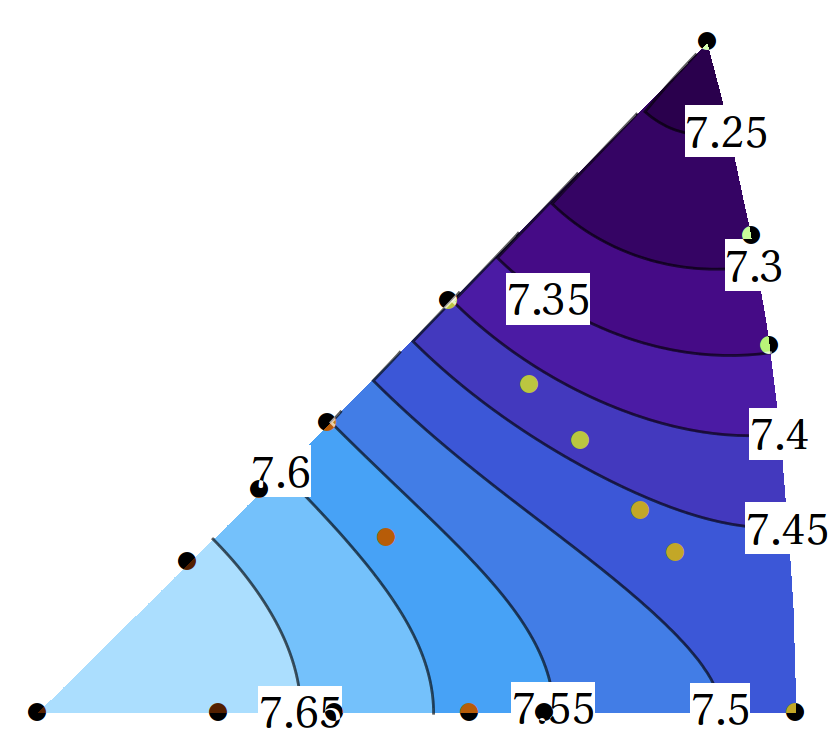}\tabularnewline
{\footnotesize (c)} & {\footnotesize (d)}\tabularnewline
\includegraphics[scale=0.25]{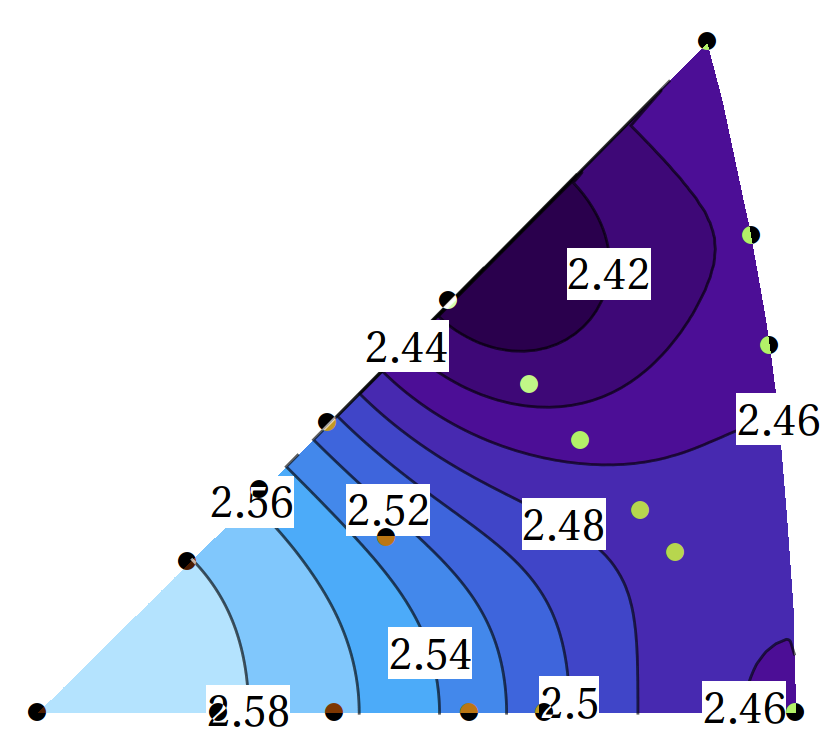} & \includegraphics[scale=0.25]{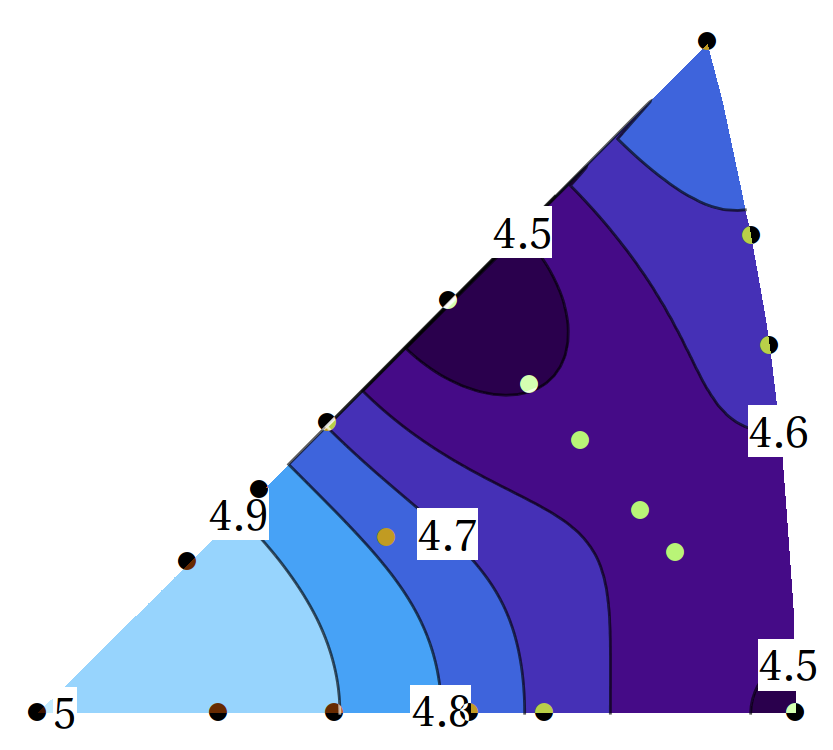}\tabularnewline
{\footnotesize (e)} & {\footnotesize (f)}\tabularnewline
\includegraphics[scale=0.25]{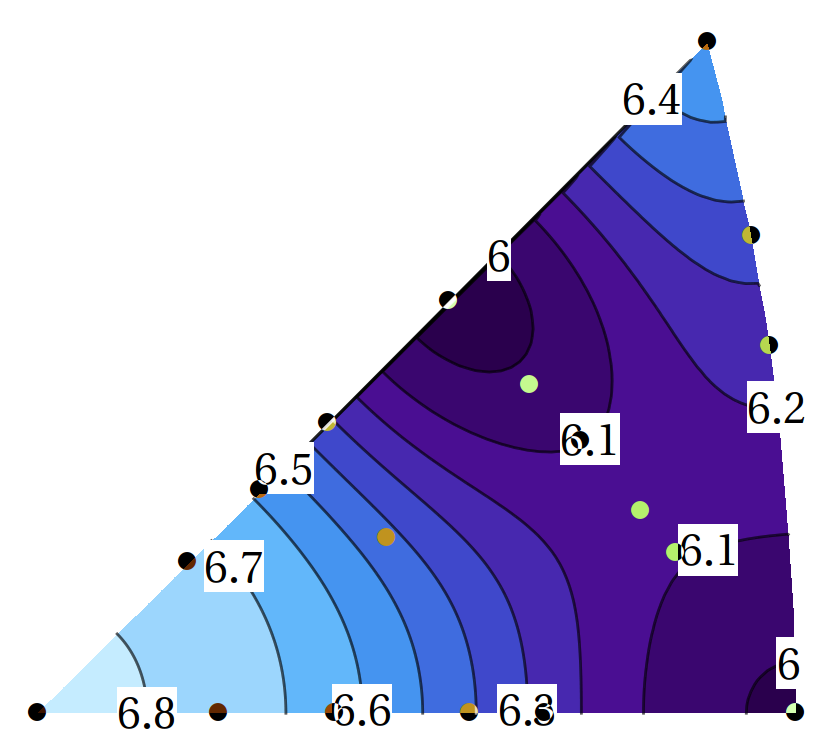} & \includegraphics[scale=0.25]{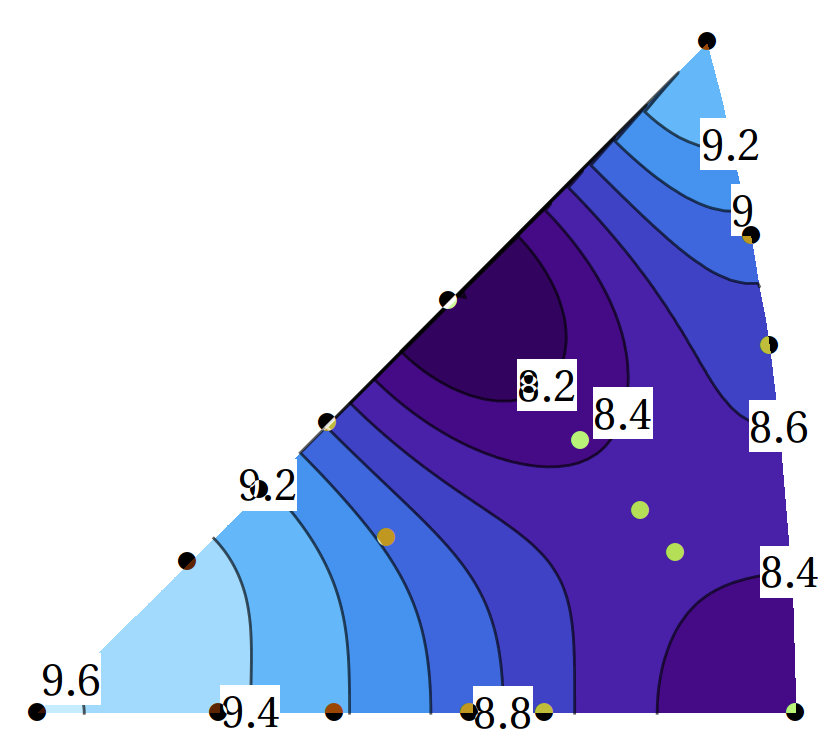}\tabularnewline
\end{tabular}\hfill{}

\caption{\label{fig:stereo}Stereographic plots of the orientation dependence
of the bcc-liquid interface free energy as a function of reduced temperature.
The contour lines characterize the fitted Kubic harmonic expression.
The 18 directions for which the interface free energy has been evaluated
are shown in panel (a) and as inverted color dots in the other panels.
Dimensionless $\gamma_{\mathrm{hkl}}$ values are shown, which has
to be multiplied by the following factors: (b): $10^{-4}$, (c)-(f):
$10^{-3}$. For panels (b) to (f), the reduced temperatures are $\epsilon=0.1$,
$0.2$, $0.3$, $0.3748$ and $0.5$, respectively.}
\end{figure*}

\section{The PFC model}

In the PFC approach, the local state of matter is characterized by
the time-averaged particle density, $\rho$. The dimensionless free
energy of the inhomogeneous (crystal + fluid) system taken relative
to a homogeneous reference fluid (of density $\rho_{\mathrm{L,ref}}$)
reads as:

\begin{equation}
\Delta F\!=\!\!\int\!\!\mathrm{d}\mathbf{r}\left\{ \frac{\psi}{2}\left[-\epsilon+(1+\nabla^{2})^{2}\right]\psi+\frac{\psi^{4}}{4}\right\} \label{eq:pfc-fe}
\end{equation}
where $\psi\propto(\rho-\rho_{\mathrm{L,ref}})/\rho_{\mathrm{L,ref}}$
is the scaled density difference. The reduced temperature $\epsilon$
can be connected to physical properties such as the bulk moduli of
the fluid and crystalline phases at the reference density and temperature.
Eq.~(\ref{eq:pfc-fe}) can be deduced \citet{key-26,key-27} from
the perturbative density functional theory of Ramakrishnan and Yussouff
\citet{key-34}. The solutions that extremize the free energy functional
can be obtained by solving the respective Euler-Lagrange equation
(ELE) \citet{key-31}:

\begin{equation}
\frac{\delta\Delta F}{\delta\psi}=\left.\frac{\delta\Delta F}{\delta\psi}\right|_{\psi_{0}}\label{eq:pfc-ele}
\end{equation}
where $\psi_{0}$ is the reduced particle density of the reference
state, while a periodic boundary condition is assumed at the borders.
Inserting the free energy functional into Eq.~(\ref{eq:pfc-ele}),
rearranging then the terms, one obtains:

\begin{equation}
[-\epsilon+(1+\nabla^{2})^{2}](\psi-\psi_{0})=-(\psi^{3}-\psi_{0}^{3})\label{eq:pfc-ele-2}
\end{equation}
The ELE has been solved here numerically, using a semi-spectral successive
approximation scheme combined with the operator-splitting method \citet{key-31,key-35}.
The computations have been performed on 33 GPU units of total theoretical
maximum computation rate of $\sim\!34.8$ TFLOPS.

The \emph{free energy of the equilibrium solid-liquid interface} has
been determined in several steps: First, the equilibrium densities
have been determined for the crystalline and liquid phases using the
ELE method \citet{key-31}. Then bcc-liquid-bcc sandwiches of equilibrium
densities and of cross-section commensurable with the actual (hkl)
face were created on which the ELE is solved using periodic boundary
conditions. Next, the solution is inserted into the expression of
grand-potential density and integrated (the contributions emerge exclusively
from the interfacial regions), dividing then the result by twice the
cross-sectional area, delivering the bcc-liquid interface free energy,
$\gamma_{hkl}$, for the hkl orientation.

\begin{figure*}
\hfill{}%
\begin{tabular}{cc}
\includegraphics[scale=0.16]{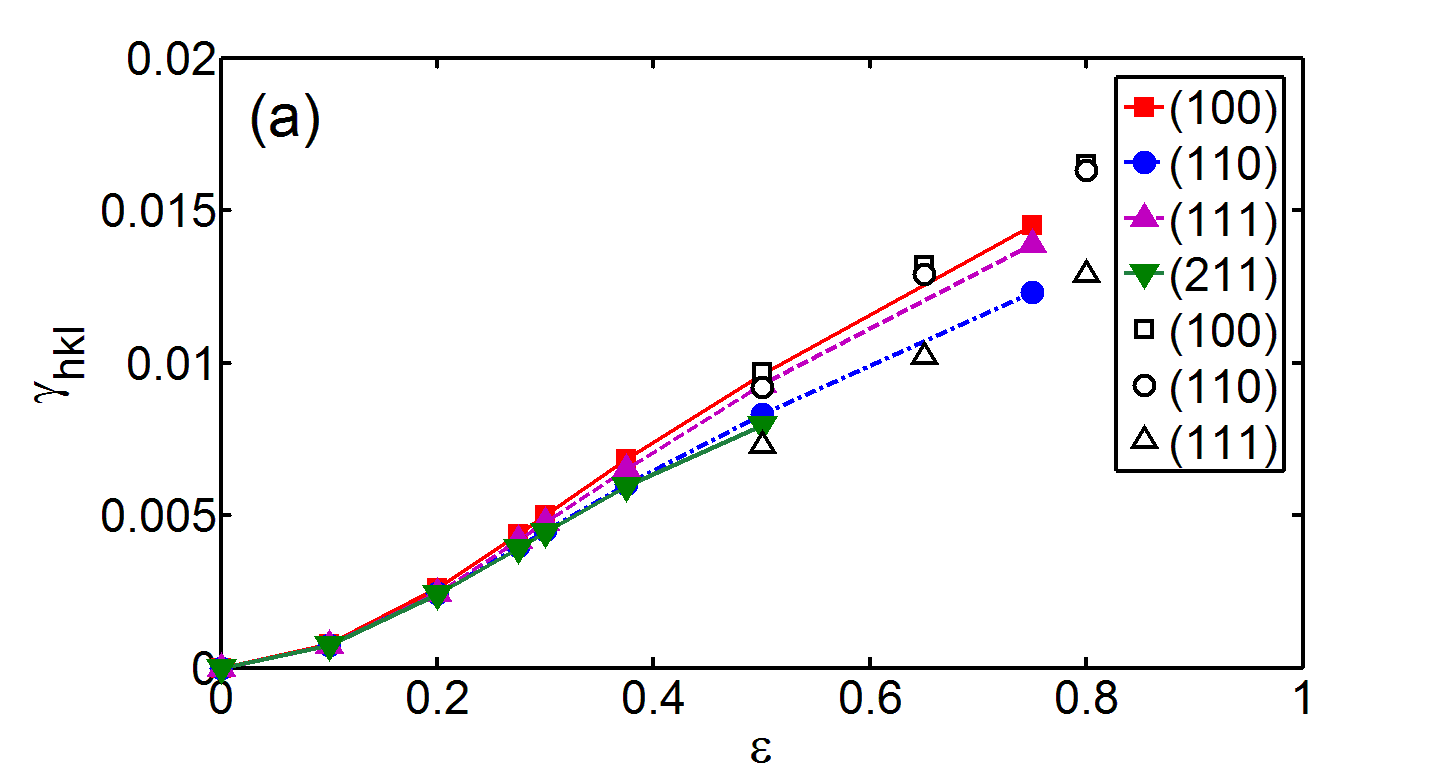} & \includegraphics[scale=0.16]{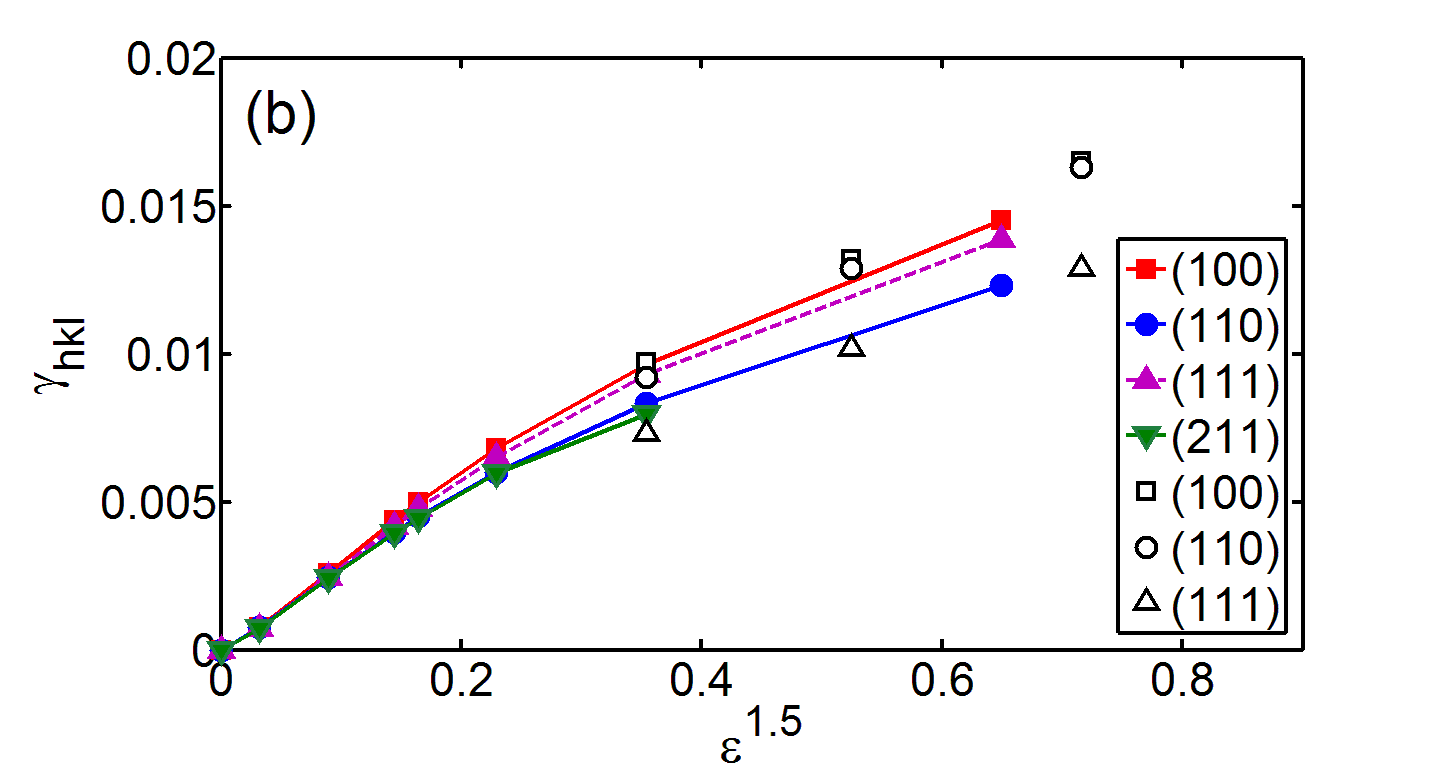}\tabularnewline
\end{tabular}\hfill{}

\caption{\label{fig:g_vs_e}(a) Bcc-liquid interface free energy (full symbols)
vs. reduced temperature for a few low index orientations as predicted
by the PFC model. (b) Interface free energy plotted vs. $\epsilon^{3/2}$.
Note that the results are consistent with the mean-field critical
exponent of the interface free energy. For comparison, data for the
fcc-liquid interface from Ref.~\citet{key-23} are also shown (empty
symbols). In contrast to the molecular dynamics results \citet{key-8,key-9,key-10},
the data for the fcc and bcc structures fall rather close to each
other.}

\end{figure*}

\section{Results and discussion}

First, we have determined $\gamma_{hkl}$ for the 18 orientations
shown in Fig.~\ref{fig:stereo}(a) for several reduced temperatures.
Next, we have fitted an eight-term expression based on the Kubic harmonic
expansion by Fehlner and Vosko \citet{key-13} (Table~\ref{tab:basis})
to the data (see the expansion coefficients and their standard deviation
in Table~\ref{tab:paramtable}). 

The respective bcc-liquid interface free energy distributions are
displayed as stereographic contour plots Fig.~\ref{fig:stereo}.
Projection is applied from the north pole of the unit sphere to the
perpendicular plane, touching the sphere at the south pole, which
coincides with the 100 direction on the maps. Only the geodesic triangle
spanned by the directions 100, 110 and 111 is shown, because of symmetry
properties of the bcc structure. At low $\epsilon\,(<0.13)$ the minimum
and maximum directions are $\left\{ 111\right\} $ and $\left\{ 100\right\} $,
respectively. For larger $\epsilon$ the maximum direction remains
the same, however, the minimum direction is $\left\{ 211\right\} $.
The interface free energy for these low index directions are shown
as a function of reduced temperature in Fig.~\ref{fig:g_vs_e}\textcolor{green}{.}
In agreement with the 2D results \citet{key-31}, the bcc-liquid data
are consistent with the known mean-field critical exponent for the
interface free energy ($1.5$). It is also clear that the anisotropy
disappears as the critical point ($\epsilon=0$) is approached. For
comparison, we have plotted $\gamma_{hkl}$ (hkl = 100, 110, and 111)
for the fcc-liquid interface at $\epsilon=0.5$, $0.65$, and $0.8$
from Ref.~\citet{key-23}. 

Remarkably, the results for the two crystal structures are quite close.
This is in accord with previous findings based on PFC nucleation data
at $\epsilon=0.3748$ \citet{key-31}, and with predictions from the
nearest-neighbor broken-bond theory \citet{key-14,key-15}. However,
the closeness of the fcc-liquid and bcc-liquid interface free energies
contradicts molecular dynamics simulations, which find that the bcc-liquid
interface free energy is about $30\%$ smaller than the data for the
fcc-liquid interface \citet{key-8,key-9,key-10}. A possible resolution
of this contradiction could be that the MD simulations refer to metallic
systems that would correspond to small $\epsilon$ values in the PFC
model (where the interface is diffuse), whereas the PFC data of Ref.~\citet{key-23}
for the fcc structure refer to very high $\epsilon$ values (where
the interface is sharp and faceted, as in the case of the nearest-neighbor
broken-bond model).

\begin{figure*}
\hfill{}%
\begin{tabular}{lll}
{\footnotesize (a)} & {\footnotesize (b)} & {\footnotesize (c)}\tabularnewline
\includegraphics[scale=0.2]{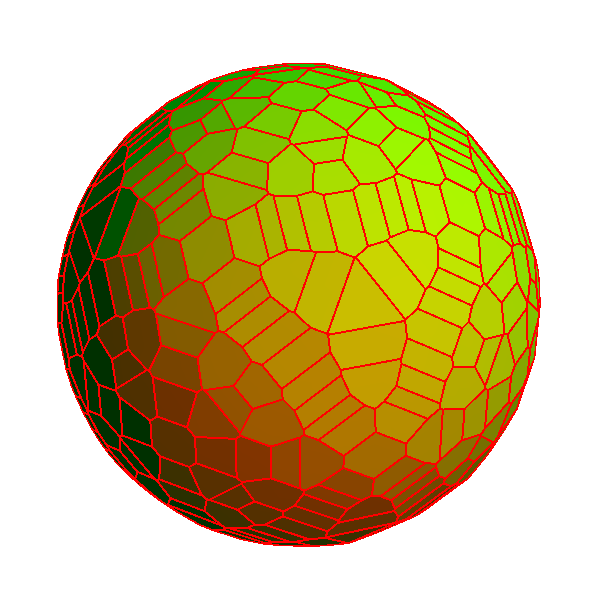} & \includegraphics[scale=0.2]{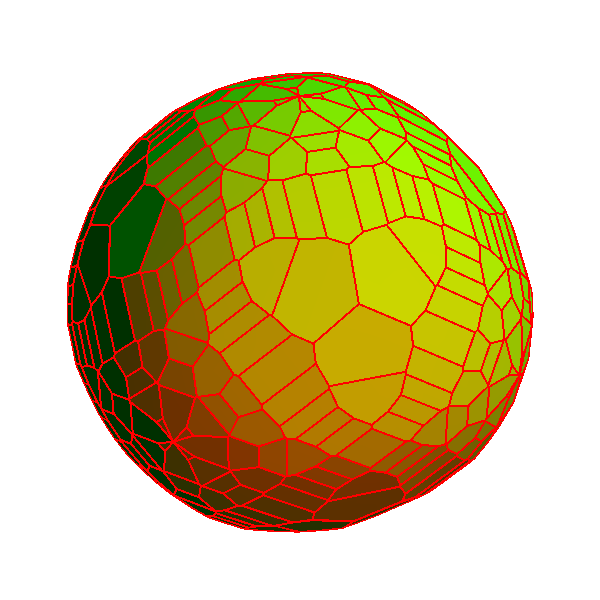} & \includegraphics[scale=0.2]{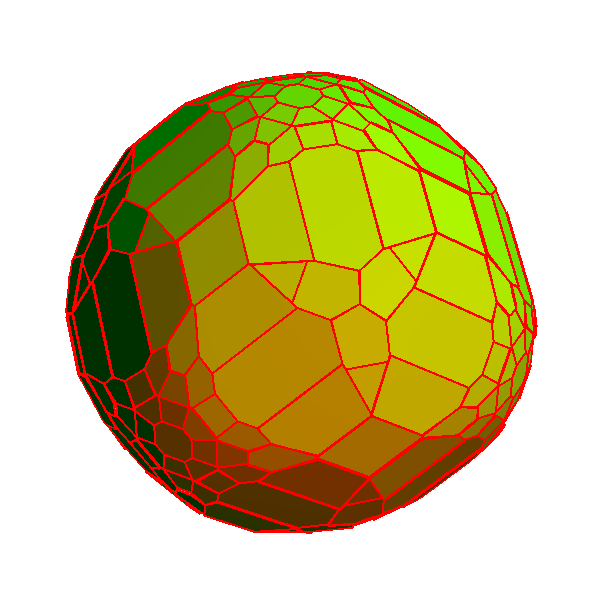}\tabularnewline
{\footnotesize (d)} & {\footnotesize (e)} & {\footnotesize (f)}\tabularnewline
\includegraphics[scale=0.2]{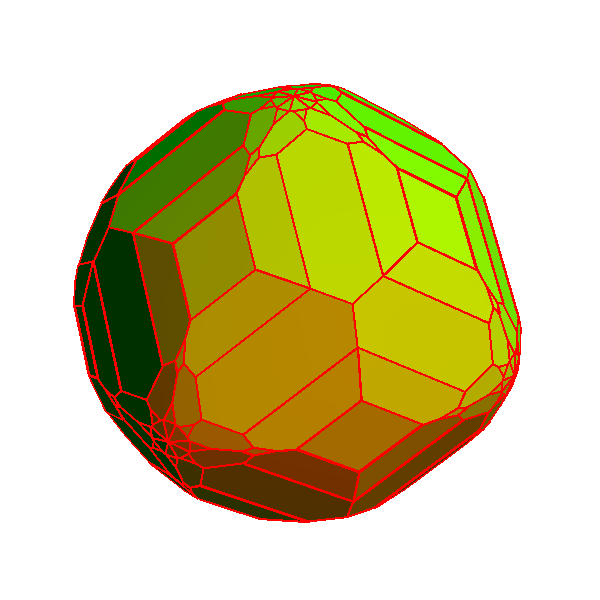} & \includegraphics[scale=0.2]{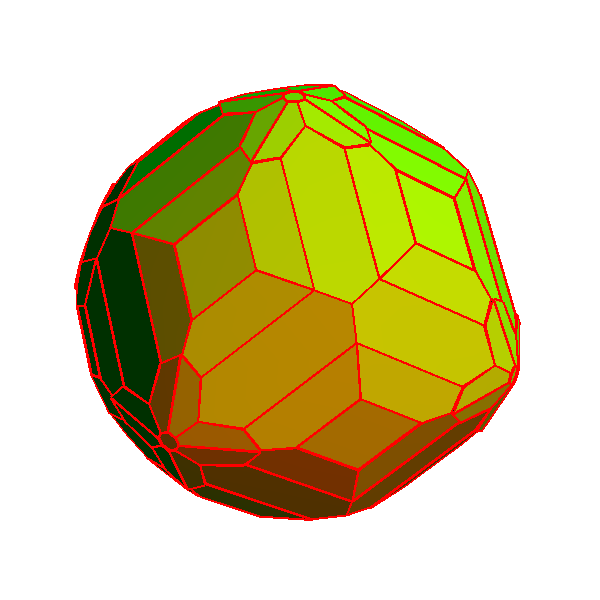} & \includegraphics[scale=0.2]{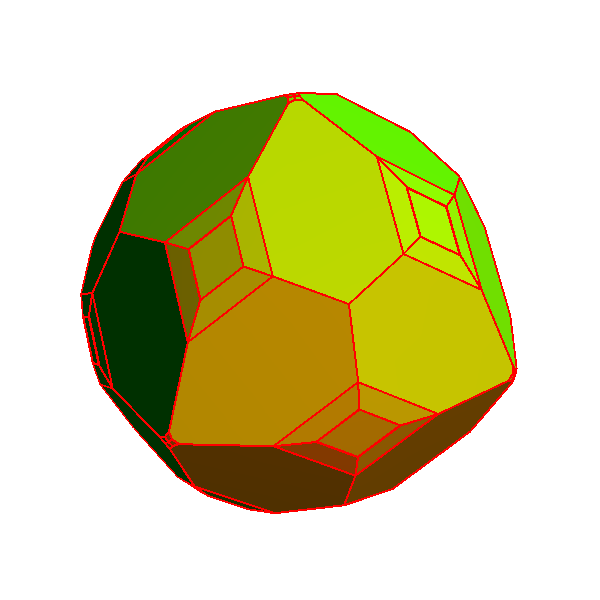}\tabularnewline
\end{tabular}\hfill{}

\caption{\label{fig:wulff_crude}Approximate Wulff shapes evaluated directly
from the 18 $\gamma_{hkl}$ data computed at the reduced temperatures
$\epsilon=0.0$, $0.1$, $0.2$, $0.3$, $0.3748$, and $0.5$, corresponding
to panels (a) to (f), respectively.}
\end{figure*}

In approximating the Wulff shapes, first we have evaluated it directly
from the 18 $\gamma_{hkl}$ values (see Fig.~\ref{fig:wulff_crude}).
Accordingly, all physical information available for the system is
accurately incorporated into the Wulff construction. We note that
a small number of $\gamma_{hkl}$ might be sufficient to describe
a Wulff shape fully if the anisotropy is high (i.e. in the case of
faceted crystals corresponding to high $\epsilon$). However, the
results are visually much less appealing if the anisotropy is low
(non-faceted crystals occurring at low $\epsilon$), since even the
surfaces that should be smooth and rounded are covered by polygons
{[}see e.g. Fig.~\ref{fig:wulff_crude}(a), where a sphere is expected{]}.
This is so, even if we use $\gamma_{hkl}$ data for 18 independent
orientations, a set which is far more numerous than usually obtained
from experiments or molecular dynamics simulations.

This problem does not occur if we use the Kubic harmonic expression,
as for any direction we may deduce $\gamma_{hkl}$. This comes, however,
at a price. The $\gamma_{hkl}$ data obtained from the fitted Kubic
harmonic expression relies on eight expansion coefficients $C_{j,k}$
obtained from a fitting procedure, which are thus subject to errors.
These errors are small if the anisotropy is low, but they may be more
significant if the anisotropy is high. Cusps in the polar plot of
the interface free energy ($\gamma$-plot) might be especially problematic.
These are the most important orientations, since ofthen the $\gamma$
value in these directions determine the whole Wulff shape. Unfortunetely,
the fit is expected to be the least accurate in the deepest cusps.

\begin{figure*}
\hfill{}%
\begin{tabular}{lll}
{\footnotesize (a)} & {\footnotesize (b)} & {\footnotesize (c)}\tabularnewline
\includegraphics[scale=0.2]{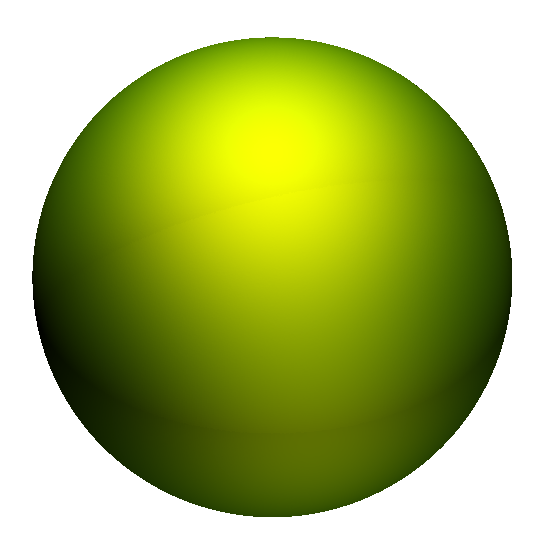} & \includegraphics[scale=0.2]{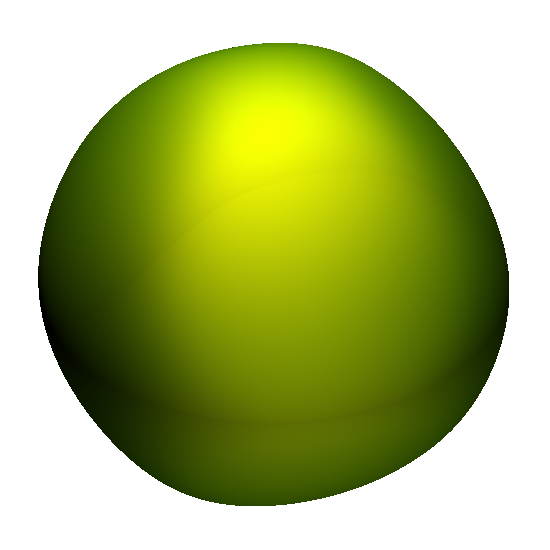} & \includegraphics[scale=0.2]{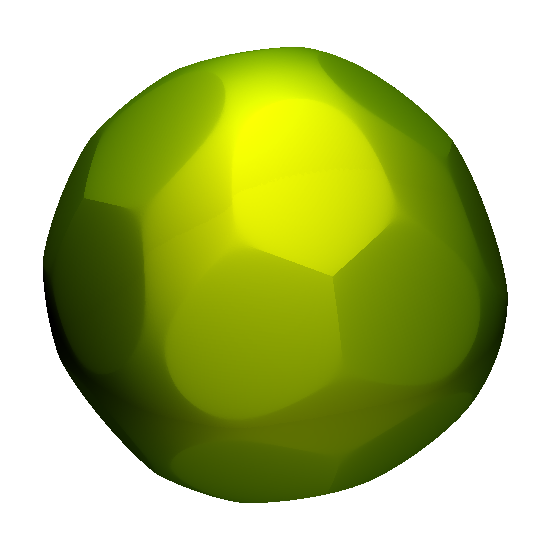}\tabularnewline
{\footnotesize (d)} & {\footnotesize (e)} & {\footnotesize (f)}\tabularnewline
\includegraphics[scale=0.2]{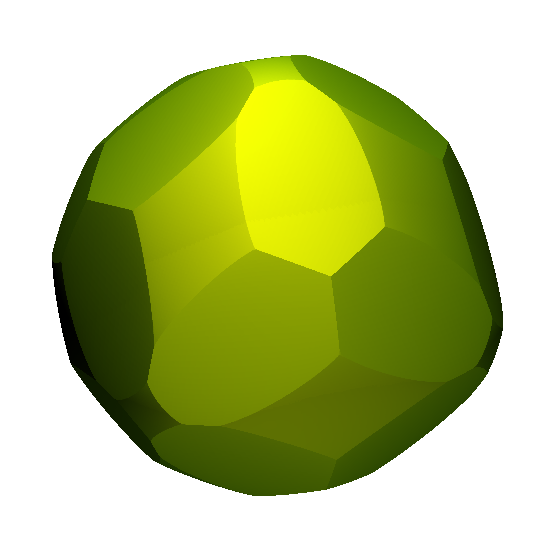} & \includegraphics[scale=0.2]{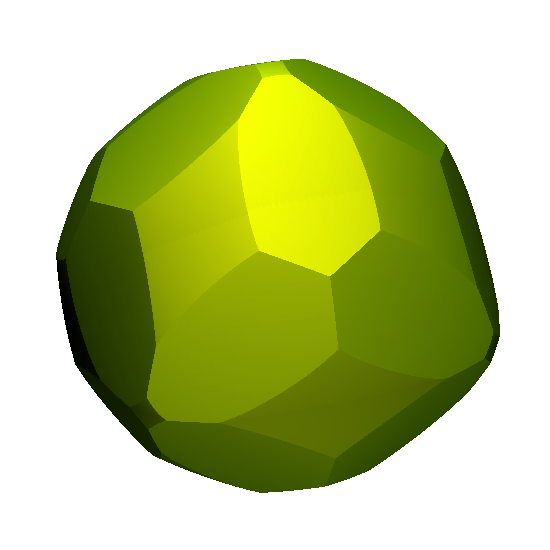} & \includegraphics[scale=0.2]{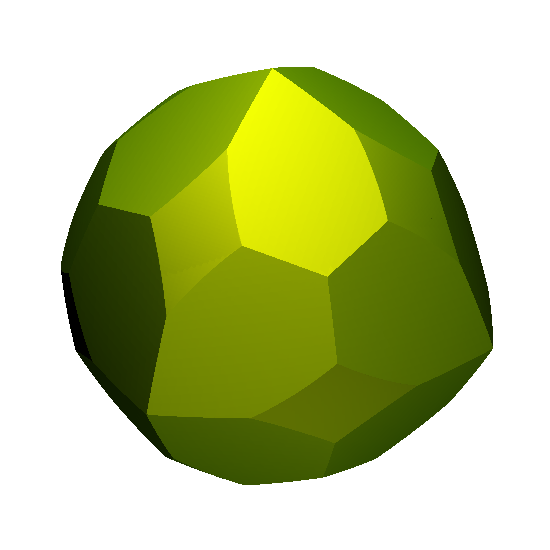}\tabularnewline
\end{tabular}\hfill{}

\caption{\label{fig:smooth_wulff}Approximate Wulff shapes computed from the
Kubic harmonic expressions (for the coefficients see Table~\ref{tab:paramtable})
that were fitted to the 18 $\gamma_{\mathrm{hkl}}$ values evaluated
at each reduced temperature ($\epsilon=0.0$, $0.1$, $0.2$, $0.3$,
$0.3748$, and $0.5$ for panels (a) to (f), respectively).}
\end{figure*}

The approximate Wulff shapes emerging from this approach are displayed
in Fig.~\ref{fig:smooth_wulff}. Indeed, the behavior for small $\epsilon$
is much improved, and we still obtain a fair reproduction for large
anisotropies. This is supported by the fact that only small changes
are seen if we merge the dense cover of orientations from the fitted
Kubic harmonic expression with the originally computed $\gamma_{hkl}$
data. However, still there are differences, which indicate that a
larger number of $\gamma_{hkl}$ is needed to have a faithful representation
of the orientation dependence of the interface free energy.

The Wulff shape changes from sphere at $\epsilon=0$ to a polyhedral
shape at large $\epsilon$ values. The latter shape differs considerably
from the rhombo-dodecahedral form found by solving the PFC equation
of motion for the growth of a bcc seed until reaching equilibrium
with the remaining liquid \citet{key-31,key-36}. The rhombo-dodecahedral
shape obtained so (which has also been predicted by the nearest-neighbor
broken-bond model as the Wulff shape for the bcc structure \citet{key-14,key-17,key-33})
might be either the result of finite size effect or simply a growth
form. Interestingly, the Wulff shapes found at large $\epsilon$ do
not resemble the shapes observed experimentally for $\mathrm{^{3}He}$
\citet{key-37} or the equilibrium crystal shapes in vapor for metals
\citet{key-38}. In our study, the dominant faces are $\left\{ 110\right\} $
and $\left\{ 211\right\} $. Apparently, the $\left\{ 100\right\} $
faces have decreasing importance with increasing $\epsilon$, whereas
the $\left\{ 111\right\} $ face does not show up at all. These differences
in the Wulff shape may be attributed to the fact that the Wulff shape
is known to be sensitive to the interaction potential \citet{key-18,key-19,key-20}.
A recent work suggests that in the PFC model the effective pair-potential
has a repulsive peak at $\approx r_{0}\sqrt{2}$, where $r_{0}$ is
the radius corresponding to the main minimum of the potential \citet{key-39},
which potential differs from the potentials expected for noble gases
and metals. Further work is needed, however, to clarify whether the
present Wulff shapes are indeed consistent with such (Dzugutov-type
\citet{key-40}) interaction potential.

\begin{table*}
\caption{\label{tab:basis}Terms $K_{j,k}$ of the Kubic harmonic expansion
with a normalization by Fehlner and Vosko \citet{key-13}.}

\hfill{}%
\begin{tabular}{l}
\hline 
$K_{0,0}=1$\tabularnewline
$K_{4,1}=\frac{1}{4}\sqrt{21}\left[5Q-3\right]$\tabularnewline
$K_{6,1}=\frac{1}{8}\sqrt{\frac{13}{2}}\left[462S+21Q-17\right]$\tabularnewline
$K_{8,2}=\frac{1}{32}\sqrt{561}\left[65Q^{2}-208S-94Q+33\right]$\tabularnewline
$K_{10,2}=\frac{1}{64}\sqrt{\frac{455}{2}}\left[7106QS+187Q^{2}-3190S-264Q+85\right]$\tabularnewline
$K_{12,2}=\frac{3}{128}\sqrt{\frac{11}{41}}\left[2704156S^{2}+352716QS+4199Q^{2}-232492S-6526Q+2423\right]$\tabularnewline
$K_{12,3}=\frac{1}{128}\sqrt{\frac{676039}{246}}\left[1025Q^{3}-16212S^{2}-8532QS-2298Q^{2}+4884S+1677Q-396\right]$\tabularnewline
$\begin{aligned}K_{14,3}=\frac{15}{256}\sqrt{51765} & \left[1311Q^{2}S+(437/18)Q^{3}-(6992/3)S^{2}-(7866/5)QS-(1577/30)Q^{2}+(1501/3)S\right.\\
 & \left.+(1109/30)Q-17/2\right]
\end{aligned}
$\tabularnewline
\tabularnewline
$Q=n_{x}^{4}+n_{y}^{4}+n_{z}^{4}\textrm{ , }S=n_{x}^{2}n_{y}^{2}n_{z}^{2}\textrm{ , }\gamma(\mathbf{n})=\sum C_{j,k}K_{j,k}\textrm{ and }\mathbf{n}=(n_{x},n_{y},n_{z})\textrm{ is the unit direction vector}$\tabularnewline
\hline 
\end{tabular}\hfill{}

\end{table*}

\begin{table*}
\caption{\label{tab:paramtable}Coefficients for the eight-term Kubic harmonic
expansion normalized by the best fit isotropic contribution $\gamma_{0}$($=C_{0,0}$),
together with their standard deviation.}

\raggedright{}\hfill{}%
\begin{tabular}{|r|rrrrrr|}
\hline 
$\epsilon$ & {\small 0.1 } & {\small 0.2 } & {\small 0.275 } & {\small 0.3 } & {\small 0.3748 } & {\small 0.5 }\tabularnewline
\hline 
$\gamma_{0}$ & {\footnotesize 7.494735E-04} & {\footnotesize 2.491726E-03} & {\footnotesize 4.126826E-03} & {\footnotesize 4.682797E-03} & {\footnotesize 6.308329E-03} & {\footnotesize 8.746856E-03}\tabularnewline
 & {\footnotesize \textpm{}4.40E-08} & {\footnotesize \textpm{}1.16E-06} & {\footnotesize \textpm{}5.59E-06} & {\footnotesize \textpm{}7.49E-06} & {\footnotesize \textpm{}1.44E-05} & {\footnotesize \textpm{}2.87E-05}\tabularnewline
\hline 
$C_{4,1}/\gamma_{0}$ & {\footnotesize 1.537295E-02} & {\footnotesize 2.065621E-02} & {\footnotesize 2.671139E-02} & {\footnotesize 2.796689E-02} & {\footnotesize 3.080253E-02} & {\footnotesize 3.498676E-02}\tabularnewline
 & {\footnotesize \textpm{}5.37E-05} & {\footnotesize \textpm{}4.24E-04} & {\footnotesize \textpm{}1.24E-03} & {\footnotesize \textpm{}1.46E-03} & {\footnotesize \textpm{}2.08E-03} & {\footnotesize \textpm{}3.00E-03}\tabularnewline
\hline 
$C_{6,1}/\gamma_{0}$ & {\footnotesize -4.906162E-03} & {\footnotesize -1.134829E-03} & {\footnotesize 7.567150E-03} & {\footnotesize 1.020313E-02} & {\footnotesize 1.508315E-02} & {\footnotesize 1.792860E-02}\tabularnewline
 & {\footnotesize \textpm{}5.96E-05} & {\footnotesize \textpm{}4.71E-04} & {\footnotesize \textpm{}1.37E-03} & {\footnotesize \textpm{}1.62E-03} & {\footnotesize \textpm{}2.31E-03} & {\footnotesize \textpm{}3.33E-03}\tabularnewline
\hline 
$C_{8,2}/\gamma_{0}$ & {\footnotesize -5.529618E-04} & {\footnotesize 1.935123E-03} & {\footnotesize 4.732360E-03} & {\footnotesize 5.636671E-03} & {\footnotesize 7.298526E-03} & {\footnotesize 1.166555E-02}\tabularnewline
 & {\footnotesize \textpm{}5.38E-05} & {\footnotesize \textpm{}4.25E-04} & {\footnotesize \textpm{}1.24E-03} & {\footnotesize \textpm{}1.47E-03} & {\footnotesize \textpm{}2.08E-03} & {\footnotesize \textpm{}3.00E-03}\tabularnewline
\hline 
$C_{10,2}/\gamma_{0}$ & {\footnotesize -2.382811E-04} & {\footnotesize -4.497379E-03} & {\footnotesize -9.913456E-03} & {\footnotesize -1.124607E-02} & {\footnotesize -1.306422E-02} & {\footnotesize -1.890355E-02}\tabularnewline
 & {\footnotesize \textpm{}4.74E-05} & {\footnotesize \textpm{}3.74E-04} & {\footnotesize \textpm{}1.09E-03} & {\footnotesize \textpm{}1.29E-03} & {\footnotesize \textpm{}1.83E-03} & {\footnotesize \textpm{}2.64E-03}\tabularnewline
\hline 
$C_{12,2}/\gamma_{0}$ & {\footnotesize -7.919240E-05} & {\footnotesize -2.207940E-03} & {\footnotesize -4.144300E-03} & {\footnotesize -4.298860E-03} & {\footnotesize -4.785945E-03} & {\footnotesize -5.170197E-03}\tabularnewline
 & {\footnotesize \textpm{}5.08E-05} & {\footnotesize \textpm{}4.02E-04} & {\footnotesize \textpm{}1.17E-03} & {\footnotesize \textpm{}1.38E-03} & {\footnotesize \textpm{}1.97E-03} & {\footnotesize \textpm{}2.84E-03}\tabularnewline
\hline 
$C_{12,3}/\gamma_{0}$ & {\footnotesize -1.286230E-05} & {\footnotesize -1.497993E-03} & {\footnotesize -2.902175E-03} & {\footnotesize -3.190361E-03} & {\footnotesize -3.794352E-03} & {\footnotesize -6.855626E-03}\tabularnewline
 & {\footnotesize \textpm{}4.63E-05} & {\footnotesize \textpm{}3.66E-04} & {\footnotesize \textpm{}1.07E-03} & {\footnotesize \textpm{}1.26E-03} & {\footnotesize \textpm{}1.79E-03} & {\footnotesize \textpm{}2.58E-03}\tabularnewline
\hline 
$C_{14,3}/\gamma_{0}$ & {\footnotesize 2.182282E-04} & {\footnotesize 2.732535E-03} & {\footnotesize 4.206570E-03} & {\footnotesize 4.540158E-03} & {\footnotesize 5.262671E-03} & {\footnotesize 1.073698E-02}\tabularnewline
 & {\footnotesize \textpm{}5.05E-05} & {\footnotesize \textpm{}3.99E-04} & {\footnotesize \textpm{}1.16E-03} & {\footnotesize \textpm{}1.38E-03} & {\footnotesize \textpm{}1.96E-03} & {\footnotesize \textpm{}2.82E-03}\tabularnewline
\hline 
\end{tabular}\hfill{}
\end{table*}

\section{Summary}

We have mapped the orientation and temperature dependence of the bcc-liquid
interface free energy $\gamma_{hkl}$ in the PFC model. As in 2D,
the dependence of $\gamma_{hkl}$ on the reduced temperature is consistent
with the respective mean field exponent. A Kubic harmonic fit has
been performed to represent the anisotropy. With increasing reduced
temperature, the corresponding Wulff shape changes from sphere (at
the critical point, $\epsilon=0$) to a polyhedral shape at large
$\epsilon$ that differs considerably from rhombo-dodecahedral, observed
in dynamic simulations based on the PFC equation of motion.

\appendix

\section*{Acknowledgement}

This work has been supported by the EU FP7 Collaborative Project ``EXOMET''
(contract no. NMP-LA-2012-280421, co-funded by ESA), and by the ESA
MAP/PECS projects “MAGNEPHAS III” (ESTEC Contract No. 4000105034/11/NL/KML)
and “GRADECET” (ESTEC Contract No. 4000104330/11/NL/KML).

\end{document}